\documentclass[pdflatex,sn-mathphys-num]{sn-jnl}

\usepackage{graphicx}%
\usepackage{multirow}%
\usepackage{amsmath,amssymb,amsfonts}%
\usepackage{amsthm}%
\usepackage{mathrsfs}%
\usepackage[title]{appendix}%
\usepackage{xcolor}%
\usepackage{textcomp}%
\usepackage{manyfoot}%
\usepackage{booktabs}%
\usepackage{algorithm}%
\usepackage{algorithmicx}%
\usepackage{algpseudocode}%
\usepackage{listings}%
\usepackage{chngcntr}

\theoremstyle{thmstyleone}%
\theoremstyle{thmstyletwo}%

\theoremstyle{thmstylethree}%

\newcommand{\kms}{km s$^{-1}$}
\newcommand{\degree}{\mbox{$^{\circ}$}}

\newcommand{\arcsec}{\mbox{$^{\prime \prime}$}}

\begin{document}

\title{Direct evidence for magnetohydrodynamic disk winds driving rotating outflows in protostar HOPS 358}

\author[1]{\fnm{Chul-Hwan} \sur{Kim}}

\author*[1,2]{\fnm{Jeong-Eun} \sur{Lee}}\email{lee.jeongeun@snu.ac.kr}

\author[3,4]{\fnm{Doug} \sur{Johnstone}}

\author[5,6]{\fnm{Gregory J.} \sur{Herczeg}}

\author[7]{\fnm{Chin-Fei} \sur{Lee}}

\author[8]{\fnm{Logan} \sur{Francis}}

\author[9]{\fnm{Patrick D.} \sur{Sheehan}}

\affil*[1]{\orgdiv{Department of Physics and Astronomy}, \orgname{Seoul National University}, \orgaddress{\street{1 Gwanak-ro}, \city{Gwanak-gu}, \postcode{08826}, \state{Seoul}, \country{Republic of Korea}}}

\affil[2]{\orgdiv{SNU Astronomy Research Center}, \orgname{Seoul National University}, \orgaddress{\street{1 Gwanak-ro}, \city{Gwanak-gu}, \postcode{08826}, \state{Seoul}, \country{Republic of Korea}}}

\affil[3]{\orgdiv{NRC Herzberg Astronomy and Astrophysics}, \orgaddress{\street{5071 West Saanich Rd}, \city{Victoria}, \postcode{V9E 2E7}, \state{British Columbia}, \country{Canada}}}

\affil[4]{\orgdiv{Department of Physics and Astronomy}, \orgname{University of Victoria}, \orgaddress{\street{3800 Finnerty Rd, Elliot Building}, \city{Victoria}, \postcode{V8P 5C2}, \state{British Columbia}, \country{Canada}}}

\affil[5]{\orgdiv{Kavli Institute for Astronomy and Astrophysics}, \orgname{Peking University}, \orgaddress{\street{Yiheyuan 5}, \city{Haidian Qu}, \postcode{100871}, \state{Beijing}, \country{China}}}

\affil[6]{\orgdiv{Department of Astronomy}, \orgname{Peking University}, \orgaddress{\street{Yiheyuan 5}, \city{Haidian Qu}, \postcode{100871}, \state{Beijing}, \country{China}}}

\affil[7]{\orgdiv{Institute of Astronomy and Astrophysics}, \orgname{Academia Sinica}, \orgaddress{\street{No. 1, Sec. 4, Roosevelt Rd}, \city{Taipei}, \postcode{106216},  \country{Taiwan}}}

\affil[8]{\orgdiv{Leiden Observatory}, \orgname{Leiden University}, \orgaddress{\street{Gorlaeus Building, Einsteinweg 55}, \city{Leiden}, \postcode{2300 RA}, \state{South Holland}, \country{The Netherlands}}}

\affil[9]{\orgdiv{National Radio Astronomy Observatory}, \orgaddress{\street{520 Edgemont Rd}, \city{Charlottesville}, \postcode{22903}, \state{Virginia}, \country{USA}}}

\abstract{
Angular momentum removal is a fundamental requirement for star and planet formation, yet the mechanisms driving this process remain debated. Magnetohydrodynamic disk winds, launched along magnetic field lines from extended disk regions, offer a promising solution, particularly in regions where magnetorotational turbulence is weak. Here we present high-resolution Atacama Large Millimeter/submillimeter Array observations of the Class 0 protostar HOPS 358, revealing a rotating, nested outflow structure traced by H$_2$CO, SO, and CH$_3$OH emission. The outflow preserves the disk’s rotational sense and is aligned with the disk axis, providing direct observational evidence for a magnetically launched disk wind. From the measured kinematics, we derive a dimensionless magnetic lever arm of approximately 2.3 and constrain the wind-launching region to radii of 10–18 astronomical units within the planet-forming zone. These results demonstrate that magnetohydrodynamic disk winds operate during the deeply embedded phase, efficiently extracting angular momentum while shaping disk evolution and establishing initial conditions for planet formation.
}


\maketitle

\section{Introduction}\label{sec:intro}
Protostars gain mass through accretion of material from the surrounding prenatal cloud core that initially lands on the disk, a process that requires efficient transport of angular momentum across a wide range of disk radii. Resolving how this angular momentum is removed or redistributed remains a fundamental question in star and planet formation theory. One candidate mechanism is turbulence driven by the magnetorotational instability (MRI), which redistributes angular momentum through coupling between magnetic fields and partially ionized gas in the disk \citep{Balbus1991}. However, detailed simulations and recent observational constraints indicate that MRI is largely inefficient within the planet-forming regions (1--30~astronomical units; au) due to low ionization levels and limited coupling \citep{Pascucci2023}. 

An alternative and increasingly favored mechanism involves radially extended magnetohydrodynamic (MHD) disk winds, which remove angular momentum by ejecting material from the disk surface along magnetic field lines \citep{Lesur2023, Pascucci2023}. These winds allow the gas remaining in the disk to drift inward while maintaining a relatively calm midplane that promotes dust settling and grain growth \citep{Taki2021, Kawasaki2025}. Recent MHD simulations and observations support that such winds can carry a significant fraction of the angular momentum budget, playing a dominant role in regulating accretion in young disks \citep{Kadam2025}. Notably, these winds are launched over a broad range of disk radii and therefore develop a nested outflow structure with flow velocities increasing closer to the protostar \citep{Pascucci2025}. A defining signature of MHD disk winds is that the outflowing gas retains the rotation direction of the disk itself, a feature that has been detected in a few protostellar systems but remains observationally challenging \citep{Bacciotti2005, Bjerkeli2016, lee_cf2018_SO_HH212, Tsukamoto2023, Omura2024, Bacciotti2025}.

Photoevaporative disk winds, driven by high-energy stellar radiation, can operate alongside MHD winds, especially at later evolutionary stages, and are believed to contribute mainly to disk dispersal rather than sustained accretion \citep{shu1993, Bai2016, Pascucci2023, Lesur2023}. Compared to MHD winds, photoevaporative winds generally show weaker rotational signatures and shallower velocity gradients in the outflowing material \citep{Hu2024}. Because dense envelopes and powerful outflows in young protostars can absorb high-energy photons before they reach the disk surface, photoevaporative winds are expected to play a more limited role during the earliest embedded phases \citep{Panoglou2012, Pascucci2023, Wilhelm2023}. Determining when and where each mechanism dominates remains crucial for understanding the initial conditions for planet formation and migration.

HOPS~358 offers a unique laboratory to test these ideas. This deeply embedded protostar has shown variable accretion activity \citep{sheehan2025}, which may modulate the disk’s physical conditions and influence wind launching. Its edge-on inclination (85$^{+2}_{-2}$\degree; \citep{Jhan2022}) provides an ideal geometry for detecting rotational signatures along the disk plane. Located in Orion~B at a distance of 404~pc \citep{Megeath2012, Dutta2024}, HOPS~358 is classified as a Class~0 protostar and was previously identified as a PACS Bright Red Source (PBRS; \citep{Stutz2013}), indicating youth, active mass infall and outflows \citep{Furlan2016, Nagy2020, Dutta2020, Jhan2022}. Accretion and outflow activities are strongly linked \citep{Watson2016}, and young embedded protostars, such as HOPS~358, often drive powerful outflows that trace the mass-loss process (e.g., \citep{Dishoeck2025}). Its outflows and jets have been observed in CO and SiO lines, and there is evidence for an episodic nature, such as knot features in the jet \citep{Dutta2020, Jhan2022, Dutta2024}. Clear rotational signatures, however, remain undetected due to previous spatial resolution limits. High-resolution observations that can directly trace the rotation within the outflow are therefore essential to reveal the role of MHD winds in shaping accretion and disk evolution in this variable, deeply embedded protostar. 

In this work, we present high-resolution Atacama Large Millimeter/submillimeter Array (ALMA) observations of HOPS~358 that directly resolve the rotational structure of the molecular outflow. We quantify the outflow morphology and kinematic properties and compare them with predictions from steady self-similar MHD disk wind models. We constrain the magnetic lever arm parameters of the outflow and launching radii. These results provide direct observational evidence that an MHD disk wind drives the observed rotating outflow in this deeply embedded protostar.

\section{Results}\label{sec:result}

To investigate the origin and properties of the outflow in HOPS~358, we analyzed several molecular lines known to be outflow tracers using high-resolution ALMA observations (see Methods, subsection~ALMA observation for details). Figure~\ref{fig:fig1} shows the integrated intensity (moment 0) and intensity-weighted velocity (moment 1) maps for multiple molecules, including disk tracers ($^{13}$CO and C$^{18}$O) and outflow tracers (CH$_{3}$OH, SO, and H$_{2}$CO). The gray contours in each panel indicate the location of the almost edge-on disk, as observed in millimeter dust continuum. The moment~1 map of $^{13}$CO reveals a clear rotational signature, with blue- and red-shifted regions located on the eastern and western sides of the continuum disk, respectively. This rotational velocity gradient is aligned with the disk major axis, which has a position angle (P.A.) of 253.4$^{+0.2}_{-0.1}$\degree~\citep{sheehan2025}. 
This indicates that the disk rotates with the eastern side moving toward us and the western side moving away (see moment 1 map of $^{13}$CO in (a) of Figure~\ref{fig:fig1}). Moreover, the power-law index estimated from the position–velocity (PV) diagram along the disk's major axis for $^{13}$CO ranges from 0.54 to 0.71, suggesting that the observed rotational feature is influenced by both Keplerian motion in the inner disk and rotation of the envelope in the outer disk (see Methods, subsection~Mass of the HOPS~358 protostar).
\par
The nearly edge-on orientation of HOPS~358 allows the rotational signatures of the outflow to be resolved. 
The molecular emission is most prominent in the southeast, where the velocity gradients associated with rotation are clearly observed.
The moment~1 maps of CH$_{3}$OH, SO, and H$_{2}$CO all exhibit rotation in the same direction as the disk. The rotation axes traced by these molecules align with the outflow axis \citep{Nagy2020, Jhan2022, Dutta2024}, although we note the presence of a localized velocity deviation relative to the rotation axis in the moment~1 map of H$_2$CO. These results suggest that the outflows traced by CH$_{3}$OH, SO, and H$_{2}$CO rotate coherently with the disk (see panel (b) of Figure~\ref{fig:fig1}).
\par

The shape of the transverse position-velocity (PV) diagram across the outflow axis provides further insight into the outflow kinematics \citep{Hirota2017}. In particular, a tilted elliptical pattern in the transverse PV diagram can be explained by a rotating, expanding thin-shell outflow model (see Supplementary Figure~1(d) of Hirota et al. \citep{Hirota2017}). Similar tilted elliptical structures have been reported in several protostellar systems, where detailed analyses consistently favor an MHD disk wind origin for these features
\citep[e.g.,][]{Hirota2017, Louvet2018, lee_cf2018_SO_HH212, Tabone2020, Valon2022}. These observational results are further supported by theoretical MHD disk wind models, which naturally reproduce such transverse PV morphologies \citep{Tabone2017}.
Figure~\ref{fig:fig2} shows the PV diagrams of H$_{2}$CO emission perpendicular to the outflow axis, extracted from slices with a width of 0.1\arcsec, centered at vertical offsets ($z_{\text{proj}}$) of 0.05\arcsec, 0.15\arcsec, 0.25\arcsec, and 0.35\arcsec\, from the disk midplane (see Figure~\ref{fig:fig1}).
Tilted elliptical shapes are clearly visible, especially at $z_{\text{proj}} \geq 0.25$\arcsec. The transverse PV diagrams of SO and CH$_{3}$OH are shown in Supplementary Figs.~2 and 3, respectively. Similar tilted elliptical patterns are visible for both molecules, although they are less pronounced at larger $z_{\text{proj}}$ due to the lower signal-to-noise ratio (S/N). 
These results support the interpretation that the outflows traced by H$_{2}$CO, SO, and CH$_{3}$OH are manifestations of disk winds. We note that, although a localized shift of the rotation axis is present for H$_2$CO, the overall tilted elliptical morphology of the transverse PV diagrams remains unchanged, indicating that the impact of this shift on the derived outflow properties is negligible.

\par

To quantify the outflow properties, we performed elliptical fitting of the transverse PV diagrams (see Methods, subsection~Outflow properties for details). Figure~\ref{fig:fig3} shows the derived outflow properties for the three molecules based on the best-fit ellipses. The outflow radii shown in (a) increase with distance from the protostar ($z_{\text{proj}}$) for all tracers. Using a parabolic fit to the outflow radii, we estimate the geometric outflow radius at $z_{\text{proj}}$ = 0, $R_{0, \mathrm{geom}}$, to be 96.0 $\pm$ 3.4 au for H$_2$CO, 76.1 $\pm$ 4.1 au for CH$_3$OH, and 34.8 $\pm$ 2.3 au for SO. This trend indicates that the emission traced by SO is located closer to the outflow axis, while H$_2$CO extends to larger radii, with CH$_3$OH distributed at intermediate radii.
\par
The opening angles are 7.2$^{+1.4}_{-1.4}$\degree, 4.8$^{+4.2}_{-4.2}$\degree, and 30.2$^{+1.1}_{-1.1}$\degree\, for H$_2$CO, CH$_3$OH, and SO, respectively. While the opening angles from H$_2$CO and CH$_3$OH molecules show relatively collimated structures, consistent with expectations from MHD disk wind models in which winds launched from smaller disk radii are more collimated, the SO-traced outflow exhibits a significantly wider opening angle than those traced by H$_2$CO and CH$_3$OH.
\par
Since HOPS~358 is observed nearly edge-on, the rotation velocity shown in (b) and the radial expansion velocity shown in (e) are only weakly affected by the inclination uncertainty. In contrast, the axial velocities shown in (d) and traced by all three molecules are strongly sensitive to the inclination, as illustrated by the range of values derived for $i=83\degree$--$87\degree$.
\par
The specific angular momentum shown in (c) traced by H$_{2}$CO slightly decreases at larger offsets from the disk midplane, while that traced by CH$_{3}$OH remains nearly constant. In contrast, the specific angular momentum traced by SO increases significantly with distance from the midplane, with a pronounced enhancement at the largest $z_{\text{proj}}$ compared to the trend observed at smaller offset ($z_{\text{proj}}$ $\leq$ 0.15\arcsec).
\par
Among the three tracers, the derived poloidal velocities (approximately 13.2 km s$^{-1}$ at $i=85\degree$) shown in (f), which describe the motion of outflowing gas along streamlines, are consistently larger than the corresponding rotational velocities (approximately 3.4 km s$^{-1}$) shown in (b). This suggests that the observed emission is unlikely to originate from the rotationally supported disk atmosphere, where the gas motion is dominated by rotation, and the poloidal component is expected to be negligible \citep{lee_cf2017_DiskAtmosphere}.

Since MHD disk wind models generally predict faster poloidal velocities for streamlines launched from smaller disk radii, poloidal velocity is often considered a key diagnostic for distinguishing wind-launching mechanisms. However, this quantity is strongly affected by the system geometry and can therefore be difficult to interpret in nearly edge-on systems.
Indeed, the poloidal velocities traced by H$_2$CO, SO, and CH$_3$OH do not exhibit a systematic trend consistent with such expectations. 
This indicates that poloidal velocity alone is insufficient to distinguish among different disk-wind origins, motivating the use of additional diagnostics.
In the following, we therefore examine the magnetic lever arm parameter, which provides a more direct constraint on the wind-launching mechanism.
\par

\section{Discussion}\label{sec:discuss}
To interpret the inferred wind properties and their implications, it is essential to consider the physical mechanisms responsible for launching disk winds.
Disk winds are generally classified into two types: MHD disk winds and photoevaporative disk winds.
MHD disk winds not only transport angular momentum and drive mass accretion but also redistribute solids within the disk.
Crystalline silicates, which can form in the heated inner disk ($>$900 K) during outbursts \citep{Abraham2009, Lee2026Nature}, have been detected not only in the inner disk but also in the cold outer regions and even in primitive solar system bodies such as comets \citep{Hanner1994, Hayward2000, Wooden2002, Shinnaka2018}. These findings pose a challenge to classical models, as the outer disk and comet-forming zones are too cold for in situ thermal annealing to occur. One proposed solution is that refractory materials are transported outward from the hot inner disk via MHD-driven winds \citep[e.g.,][]{shu1996, Giacalone2019}. If such transport occurs during the earliest evolutionary stages, as some models predict, MHD disk winds could directly influence the composition of planet-forming material well before disk clearing begins.
In contrast, photoevaporative disk winds primarily contribute to disk clearing and have a minimal effect on accretion \citep[e.g.,][]{Pascucci2023}. Distinguishing between these two mechanisms is, therefore, crucial for understanding how protostars grow and how material is redistributed in the star formation process.
\par

A key parameter for differentiating disk wind origins is the magnetic lever arm parameter \citep{Blandford1982, Ferreira2006, Louvet2018, Tabone2020, Nazari2024}. This parameter, denoted as $\lambda_{\text{BP}}$, is defined as the ratio of the total specific angular momentum extracted by the wind to the initial specific angular momentum at its launching point \citep{Blandford1982}. For MHD disk winds, the magnetic lever arm is expected to exceed 1.5, while for photoevaporative disk winds, it is typically 1 \citep{Louvet2018}.
Figure~\ref{fig:fig4} shows the observed relationship between specific angular momentum and total velocity, both normalized by the square root of the protostar mass, derived from the best-fit transverse PV diagrams at each $z_{\text{proj}}$. These measurements are compared with theoretical expectations from steady self-similar cold MHD disk wind models for different magnetic lever arm values, $\lambda_{\phi}$, which provide a lower limit on $\lambda_{\text{BP}}$, as well as for different launching radii, thereby allowing representative magnetic lever arm values to be inferred (see Methods, subsection Magnetic lever arm parameter and Outer launching radius). This comparison allows the inferred magnetic lever arm to be interpreted as a measure of the efficiency of angular momentum extraction under the assumption of purely magnetic driving. 

If the wind is partially magnetothermal in nature, thermal pressure may additionally contribute to the outflow properties. In this case, the inferred magnetic lever arm should be interpreted as an effective parameter, corresponding to the value required in a cold MHD wind formalism to reproduce the observed kinematics.
The inferred magnetic lever arm parameters for the disk winds traced by H$_2$CO, SO, and CH$_3$OH all exceed 1.5 and cluster near 2.3. These values remain above 1.5 even when adopting an inclination of 83\degree, demonstrating that this result is robust against the inclination uncertainty (see Figure~\ref{fig:fig4}). Since the adopted protostar mass of 1.42 $\pm$ 0.40 $M_{\odot}$ represents an upper limit (see Methods, subsection~Mass of the HOPS~358 protostar), these parameters should be regarded as lower limits. These results provide compelling evidence that the outflows traced by these molecules in HOPS~358 are driven by an MHD disk wind.
\par

The outer launching radii of these winds are also estimated using the derived outflow properties (see Methods, subsection~Magnetic lever arm parameter and Outer launching radius). For H$_2$CO, SO, and CH$_3$OH, the outer launching radii inferred assuming an inclination of 85\degree\,are 18.0~$\pm$~5.3~au, 10.1~$\pm$~1.6~au, and 10.9~$\pm$~0.6~au, respectively. These values lie well within the disk’s extent (disk radius of approximately 71.9~au adjusted for the distance of 404~pc; \citep{sheehan2025}).
This radial stratification is also evident in the observed emission distribution. Figure~\ref{fig:fig5} shows that SO traces the inner region of the outflow, H$_{2}$CO traces the outer region, and CH$_{3}$OH occupies an intermediate region, indicating a morphologically nested outflow structure that is a characteristic feature of MHD disk winds \citep{Pascucci2025}.
Taken together, these results suggest that H$_2$CO traces an outer MHD disk wind layer launched from farther out in the disk, SO traces an inner wind layer originating closer to the protostar, and CH$_3$OH traces an intermediate wind layer. This nested morphology implies the removal of angular momentum from the disk over a broad range of radii, thereby allowing inward transport of mass.
Although HOPS~358 has an edge-on geometry, which limits a clear view of increasing outflow velocities toward the center, the inferred nested structure and the lever arm values strongly support an MHD origin. 

\par
H$_2$CO and CH$_3$OH can be released into the gas phase from ice mantles on grain surfaces in the outflow \citep{tychoniec2021}. Because the binding energy of H$_2$CO is lower than that of CH$_3$OH \citep{Penteado2017, Minissale2022}, H$_2$CO can be liberated under weaker shocks. In contrast, SO abundance is enhanced in shocks within dense regions exposed to local UV radiation fields \citep{Gelder2021, leeje2023}, which are expected to be stronger near the protostar. 
These chemical pathways suggest an explanation for the different observed launching regions traced by the three molecules, such that SO, requiring UV, should be strongest from the inner disk scales, and H$_{2}$CO should extend to the largest disk scales.
However, the observed SO emission extends to substantially larger heights than CH$_3$OH, exhibiting an increasing opening angle with altitude, indicating that additional processes beyond the launching regions must be at play.

Beyond shaping the launching regions, chemical effects, particularly UV irradiation, also influence the emission distributions along the outflow axis. Thermo-chemical modeling of magnetized protostellar disk winds by \citet{Panoglou2012} demonstrates that, in Class 0 systems, the radiation field varies systematically with height along the wind streamlines, owing to the combined effects of geometry, column density, and dust shielding. These models show that molecular survival in disk winds is governed by the competition between photodissociation, chemical formation, and the wind flow timescale, although the calculations primarily focus on species such as H$_2$ and CO.

Motivated by this framework, we compare the formation and photodissociation timescales of SO and CH$_3$OH to interpret their different vertical emission extents. Although the photodissociation rates of CH$_3$OH and SO are comparable (approximately 10$^{-9}$~s$^{-1}$; \citep{Heays2017}), their gas-phase formation pathways differ substantially. SO can be efficiently reformed through rapid gas-phase reactions at heights where the radiation field strength is below a threshold \citep{Panoglou2012}, whereas gas-phase formation of CH$_3$OH is much slower, with a characteristic timescale of approximately $10^{5}$~yr in cold dark clouds \citep{Garrod2006}. 
As a result, at larger $z_{\text{proj}}$, where UV shielding by the inner wind layers becomes less effective, UV radiation penetrates into the intermediate wind layer. In this wind layer, CH$_3$OH is efficiently photodissociated, while SO can remain detectable owing to its rapid reformation, provided that the UV intensity remains below the threshold.
In this picture, SO predominantly traces the inner wind layer at lower altitudes ($z_{\text{proj}}\,\leq\,0.15\arcsec$), while at larger heights ($z_{\text{proj}}\,\simeq\,0.25\arcsec$) its detectable emission extends into the intermediate wind layer. This interpretation is in good agreement with the significant increase in the specific angular momentum of SO at $z_{\text{proj}}\,=\,0.25\arcsec$ (see (c) in Figure~\ref{fig:fig3}), as well as the larger outer launching radius at that height inferred from comparisons between the observed outflow properties and the expected relations from MHD disk wind models (see Figure~\ref{fig:fig4}).
\par
To evaluate whether the MHD disk wind can sustain accretion, it is necessary to compare the mass-loss rate of the MHD disk wind and the mass accretion rate. We estimated the mass-loss rate, $\dot M_{\text{wind}}$, for the H$_{2}$CO wind by assuming optically thin and local thermodynamic equilibrium (LTE) conditions, taking the average of the mass-loss rates derived from each slice and multiplying by two to account for the contribution from both sides of the disk (see Methods, subsection~Calculation of disk wind mass-loss rate and stellar mass accretion rate for details).
The mass-loss rate is (2.04 $\pm$ 0.09)~$\times$~10$^{-5}$~$M_{\odot}$~yr$^{-1}$, assuming an inclination of 85\degree.
The mass accretion rate is estimated at (2.71 $\pm$ 0.76)~$\times$~10$^{-6}$~$M_{\odot}$~yr$^{-1}$, based on the physical properties of HOPS 358 derived in this work and reported in previous works \citep{Furlan2016, cfLee2020, Podio2021} (see Methods, subsection~Calculation of disk wind mass-loss rate and stellar mass accretion rate for details). Both the mass-loss rate and the mass accretion rate are considered as lower limits, due to the assumptions made in the derivation. Since the wind mass-loss rate is about 7.5 times higher than the accretion rate, the observed wind could plausibly drive the accretion flow.
\par
Jets, which are fast and highly collimated, are launched from the innermost disk region through MHD processes \citep{Shang2007, Zanni2013, frank2014}. Thus, jets and MHD disk winds are likely closely linked, and jets may arise from the innermost, fastest streamlines of an MHD disk wind. SiO emission, commonly used as a tracer of protostellar jets in the Class 0 stage \citep{cfLee2020}, is detected at larger distances from the protostar, where a chain of knotty structures indicative of a periodic variation in the jet velocity ($P$ = 22~yr), likely linked to episodic accretion, is observed \citep{Jhan2022, Dutta2024}. In contrast, we find no evidence for a high-velocity SiO component within the inner region where H$_2$CO, SO, and CH$_3$OH trace the rotating molecular outflow (Methods, subsection~ALMA observations).

If the molecular outflow were entrained by a high-velocity jet, one would expect (1) the presence of high-velocity molecular emission and/or (2) a narrow, highly collimated morphology aligned with the jet axis. Neither of these signatures is observed in any of the three molecular tracers.
Instead, the velocity structure and spatial extent of the molecular emission are naturally explained by an MHD disk wind launched over an extended range of disk radii. In this scenario, rotation is an intrinsic property of the flow rather than a by-product of jet entrainment. The absence of jet-like kinematic signatures therefore argues against entrainment and favors direct launching from the disk. We thus conclude that the rotating molecular outflows analyzed in this work are not driven by interaction with a fast jet, but instead originate as a disk wind.

Future infrared observations of atomic jet tracers with the James Webb Space Telescope (JWST) will provide a critical test of possible jet–disk wind interactions in the innermost regions \citep{Pascucci2025}. Nevertheless, the current data strongly indicate that the rotating molecular outflow component studied here is physically distinct from the SiO jet observed at larger distances.

Determining the origin of disk winds is key to understanding disk evolution and protostellar growth. The observed rotational signatures and nested outflow structures in the very young protostar HOPS 358, traced by multiple molecular species, provide strong constraints on the properties and origin of the wind. Because different molecules probe distinct excitation conditions and launching radii within the disk, this multi-line analysis enables us to characterize an MHD disk wind over a broad radial extent. This, in turn, yields a detailed picture of how the wind structure and efficiency vary with launching radius, and how angular momentum is transported across the wind-launching region of the disk during the earliest stages of star formation, when mass accretion is most vigorous. These winds may also redistribute crystalline silicates from the hot inner disk to comet-forming regions even at the extremely young Class 0 stage. Together, our results suggest that the MHD disk wind plays a crucial role in shaping protostellar systems from their earliest phases.

\clearpage

\section{Methods}\label{sec:methods}

\subsection{ALMA observations}\label{sec:ALMAobs}
HOPS 358 was observed with the Atacama Large Millimeter/submillimeter Array (ALMA) as part of program 2023.1.01245.S (PI: Chul-Hwan Kim) on 2023 October 1 and 2. The observation consists of two execution blocks (EBs) using baselines ranging from 92.1 m to 8.5 km. The spectral setup includes eight spectral windows (SPWs) in ALMA Band 6 (about 230 GHz). Seven of the SPWs cover different molecular lines known as outflow/jet and disk tracers: SiO 5-4, H$_{2}$CO 3$_{2,1}$-2$_{2,0}$, C$^{18}$O 2-1, SO 6$_{5}$-5$_{4}$, $^{13}$CO 2-1, CH$_{3}$OH 3$_{-2,2}$-4$_{-1,4}$, and CO 2-1. To explore the launching mechanism of outflows, this paper focuses on the molecular lines tracing outflowing material near the protostar, including H$_{2}$CO, SO, and CH$_{3}$OH lines. The SPWs containing H$_{2}$CO, C$^{18}$O, SO, and $^{13}$CO have a bandwidth of 0.059 GHz with a channel width of 0.122 MHz ($\delta$\textit{v}=0.17 \kms), while the SPW targeting CH$_{3}$OH has a bandwidth of 0.177 GHz with a channel width of 0.122 MHz ($\delta$\textit{v}=0.34 \kms).
One SPW is dedicated to continuum observation, with a bandwidth of 2 GHz and a spectral resolution of 15.6 MHz.

\par
The data were initially calibrated using Common Astronomy Software Applications (CASA) v6.5.4 \citep{McMullin2007}. To achieve a high S/N, we employed natural weighting in the tclean task of CASA. The continuum data were created using the dedicated continuum SPW and the continuum channels for each SPW provided by the pipeline to enhance the S/N of the continuum data. The synthesized beam of the continuum data is 0.08\arcsec\,$\times$ 0.07\arcsec. The synthesized beam for the molecular line data of H$_{2}$CO, C$^{18}$O, SO, and $^{13}$CO is 0.10\arcsec\,$\times$ 0.07\arcsec, while the synthesized beam size for the CH$_{3}$OH line data is 0.09\arcsec\,$\times$ 0.07\arcsec. 

To increase the S/N of the molecular line data, we used CASA's specsmooth task. In specsmooth, we set the function parameter to hanning. As a result, the channel width of each molecular line data is twice that mentioned above. Finally, the root-mean-square (rms) noise level of continuum data is 0.14 mJy beam$^{-1}$, while the rms noise levels of molecular line data range from 1.2 to 2.5 mJy beam$^{-1}$ 
($\sigma_{\mathrm{C^{18}O}}$ = 2.1 mJy beam$^{-1}$, 
$\sigma_{\mathrm{^{13}CO}}$ = 2.5 mJy beam$^{-1}$, 
$\sigma_{\mathrm{H_{2}CO}}$ = 1.6 mJy beam$^{-1}$, 
$\sigma_{\mathrm{SO}}$ = 2.2 mJy beam$^{-1}$,  and
$\sigma_{\mathrm{CH_{3}OH}}$ = 2.1 mJy beam$^{-1}$).

\subsection{Mass of the HOPS 358 protostar}\label{sec:mass}
$^{13}$CO clearly shows the rotational feature of the disk (see Figure~\ref{fig:fig1}), enabling us to estimate the protostar mass. We constructed a PV diagram of the $^{13}$CO emission along the disk major axis (see Supplementary Fig.~\ref{extfig:SupplyFig4}) and analyzed it using the pvanalysis module within the SLAM toolbox, a Python package for analyzing PV diagrams \citep{Aso2023}.

The pvanalysis module provides best-fit solutions using two approaches, the ridge method and the edge method. Both methods fit emission above a user-defined threshold using a single (or double) power-law function with the Markov Chain Monte Carlo method. In the edge method, the data point used for fitting is taken as the outermost position exceeding the adopted intensity threshold in the one-dimensional profile extracted from the PV diagram. In contrast, the ridge method defines it as the intensity-weighted mean position within the same profile \citep{Aso2023}.
We adopted a threshold of 4$\sigma_{\mathrm{^{13}CO}}$ and applied a single power-law function in both cases. Because the uncertainties reported by SLAM account only for statistical errors, we further explored the effects of uncertainties in the disk inclination and P.A. on the derived parameters (see Supplementary Table~\ref{tab:protostar_mass}).

The derived power-law index and central mass are not significantly affected by the uncertainties in the disk inclination and P.A. Accounting for these effects, the ridge method yields a power-law index of 0.54 $\pm$ 0.003 and a central mass of 1.02 $\pm$ 0.01 $M_{\odot}$, while the edge method gives values of 0.71 $\pm$ 0.007 and 1.81 $\pm$ 0.01 $M_{\odot}$, respectively.
The power-law index obtained from the ridge method is almost consistent with the Keplerian motion, which requires an index of 0.5. For the edge method, the power-law index ranges between 0.5 and 1. This suggests that the observed rotation feature of the $^{13}$CO emission reflects the combined effects of the infalling, rotating envelope and the Keplerian disk. 
\par
Previous work by \citet{Aso2023} reported that the ridge method tends to underestimate the central mass by approximately 30\%, whereas the edge method can overestimate it by up to a factor of two. Therefore, we adopt the mean of the two values as our best estimate of the protostar mass, with the uncertainty given by the standard deviation, which exceeds the propagated statistical error. This yields a final protostar mass of 1.42 $\pm$ 0.40 $M_{\odot}$.
We note that, although this average provides a conservative estimate, the lower mass derived from the ridge method, which traces emission closer to the inner disk and yields the power-law index consistent with Keplerian rotation, may be closer to the true protostar mass. In this sense, the adopted average value can be regarded as an upper limit on the central mass.

The C$^{18}$O emission also distributes along the disk structure in HOPS 358. While the best-fit PV diagram of C$^{18}$O yields a protostellar mass (1.41 $\pm$ 0.11~$M_{\odot}$) consistent with that from $^{13}$CO, the derived power-law indices (0.909 from the ridge method and 1.066 from the edge method) are both close to unity, suggesting that the C$^{18}$O emission arises from the infalling rotating envelope rather than from the Keplerian disk. We therefore use only the $^{13}$CO data to estimate the protostellar mass.

\subsection{Outflow properties}\label{sec:outflow_prop}
We analyzed the transverse PV diagrams at different offset positions, z$_{\text{proj}}$, from the disk mid-plane to derive the physical properties of outflow, including the outflow radius, $R_{\text{outflow}}$, rotation velocity, $v_{\phi}$, axial velocity, $v_{\text{z}}$, radial expansion velocity, $v_{\text{r}}$, poloidal (or outflowing) velocity, $v_{\text{p}}$, and specific angular momentum, $j$. Each transverse PV diagram was fitted with an ellipse, assuming a rotating and expanding thin-shell outflow structure \citep{Hirota2017}. Only emission above 2$\sigma_{\text{mole}}$ was considered in the fitting, focusing on the outer part of each PV ellipse. The ellipse fittings were performed using the random sample consensus (RANSAC) algorithm implemented in scikit-image \citep{skimage}, which robustly determines model parameters by iteratively fitting random subsets of the data and rejecting outliers. We adopted hyperparameters of min\_samples = 5, residual\_threshold = 5, and max\_trial = 100. To estimate the uncertainties of the elliptical fitting results, we applied a bootstrap method by performing 1000 RANSAC ellipse fittings for each transverse PV diagram. The mean and standard deviation of the resulting 1000 fits were adopted as the best-fit value and corresponding uncertainty for each transverse PV diagram.

Figure~\ref{fig:fig2} shows the best-fit ellipse as a colored dashed ellipse for each PV diagram. The purple dots on the ellipse represent gas parcels located at $\theta$ = 0 and $\pi$, while the green dots correspond to $\theta$ = $\pm\frac{\pi}{2}$ (see Supplementary Fig.~\ref{extfig:SupplyFig1}). At $\theta$ = 0 and $\pi$, the observed radial velocities ($v_{\theta=0}$ and $v_{\theta=\pi}$) result from a combination of the axial velocity ($v_{\text{z}}$) and radial expansion velocity ($v_{\text{r}}$). In contrast, at $\theta$ = $\frac{\pi}{2}$ and $-\frac{\pi}{2}$, the radial velocities ($v_{\theta=\pm\frac{\pi}{2}}$) arise from a combination of $v_{\text{z}}$ and rotation velocity ($v_{\phi}$), and half of their observed spatial offsets ($r_{\theta=\pm\frac{\pi}{2}}$) corresponds to the radius of the outflow shell from the center ($R_{\text{outflow}}$). In the convention, blueshifted velocities are defined as negative, and redshifted velocities are defined as positive. The spatial offsets associated with the blueshifted and redshifted components are defined to be positive and negative, respectively. 
The uncertainties in the derived parameters are estimated from the uncertainties in the best-fit results of elliptical fitting.

The outflow properties are derived using the parameters, $v_{\theta=0}$, $v_{\theta=\pi}$, $v_{\theta=\pm\frac{\pi}{2}}$, and $r_{\theta=\pm\frac{\pi}{2}}$, as follows:
\begin{align}\label{eq:R_out}
 R_{\text{outflow}} = \frac{|r_{\theta=\frac{\pi}{2}}|+|r_{\theta=-\frac{\pi}{2}}|}{2},
\end{align}
\begin{align}\label{eq:Vphi}
 \textit{v}_{\phi} = \frac{v_{\theta=-\frac{\pi}{2}}-v_{\theta=\frac{\pi}{2}}}{2\text{sin}\,i},
\end{align}
\begin{align}\label{eq:Vexp}
 \textit{v}_{\text{r}} =  \frac{v_{\theta=\pi}-v_{\theta=0}}{2\text{sin}\,i},
\end{align}
\begin{align}\label{eq:Vz}
 \textit{v}_{\text{z}} = -\frac{v_{\theta=-\frac{\pi}{2}}+v_{\theta=\frac{\pi}{2}}}{2\text{cos}\,i}\;\; \text{or}\;\; -\frac{v_{\theta=\pi}+v_{\theta=0}}{2\text{cos}\,i},
\end{align}
\begin{align}\label{eq:Vp}
 \textit{v}_{\text{p}} = \sqrt{\textit{v}_{\text{z}}^2 + \textit{v}_{\text{r}}^2},
\end{align}
\begin{align}\label{eq:jz}
 \textit{j} =  R_{\text{outflow}} \times \textit{v}_{\phi},
\end{align}
where \textit{i} is the inclination angle of the outflow axis from the line of sight, $85^{+2}_{-2}$\degree, $v_{\text{p}}$ is the poloidal velocity of the outflow along the streamline, and $j$ is the specific angular momentum of the outflow. The uncertainties in the outflow properties are estimated by propagating the uncertainties of the fitted parameters only. 
In addition, to assess the systematic effect of the inclination uncertainty, we also recompute the relevant outflow properties assuming inclinations of $83^\circ$ and $87^\circ$.

We also performed parabolic fitting to the measured outflow radii for each molecular tracer to estimate the geometric launching radius, $R_{0,\mathrm{geom}}$, at $z_{\text{proj}}$ = 0. The values of $R_{0,\mathrm{geom}}$ for molecules are shown as open symbols in panel (a) of Figure~\ref{fig:fig3}. This fitting allows us to characterize the opening angle of the outflow traced by each molecule. The opening angle, $\theta_{\mathrm{opening}}$, was calculated using the outflow radius measured at the largest height, $R_{\mathrm{largest}}$, and the corresponding vertical offset, $z_{\mathrm{deproj, largest}}$ in units of au, as
\begin{align}\label{eq:openingangle}
 \theta_{\mathrm{opening}} =  \arctan(\frac{R_{\mathrm{largest}} - R_{0,\mathrm{geom}}}{z_{\mathrm{deproj, largest}}}),
\end{align}
where $z_{\mathrm{deproj, largest}}$ is given by $z_{\mathrm{proj, largest}} / \sin i$.

\subsection{Magnetic lever arm parameter and Outer launching radius}\label{sec:mla_launchR}
The dimensionless magnetic lever arm, $\lambda_{\phi}$, can be expressed as:
\begin{align}\label{eq:R_out}
    (R_{\text{outflow}}\textit{v}_{\phi}\times \textit{v}_{\text{tot}})/GM_{\star}=\lambda_{\phi}\sqrt{2\lambda_{\phi}-3},
\end{align}
where $R_{\text{outflow}}$ is the outflow radius, $v_{\phi}$ is the rotation velocity, $v_{\text{tot}}$ is the total velocity of the outflowing gas ($v_{\text{tot}} = \sqrt{{v_{\phi}}^2 + v_{\text{p}}^2}$), $G$ is the gravitational constant, and $M_{\star}$ is the protostellar mass \citep{Blandford1982, Louvet2018, Nazari2024}.
The parameter $\lambda_{\phi}$ quantifies the specific angular momentum considering only the rotational motion of the matter. Therefore, $\lambda_{\phi}$ provides a lower limit for $\lambda_{\text{BP}}$, which accounts not only for matter rotation but also for magnetic torsion \citep{Blandford1982, Ferreira2006, Tabone2020}.
\par
The derived outflow properties allow us to estimate the outer launching radius, $r_{0}$, of the disk wind assuming a cold, steady, axisymmetric MHD disk wind, neglecting the gravitational potential \citep{Anderson2003}. 
The relation of \citet{Anderson2003} can be expressed as:
\begin{align}\label{eq:r0}(\textit{v}_{\text{p}}^{2}+\textit{v}_{\phi}^{2})r_0^{3/2}+3(GM_{\star})r_0^{1/2}-2R_{\text{outflow}}\textit{v}_{\phi}(GM_{\star})^{1/2} \approx 0,
\end{align}
where $v_{\text{p}}$ is the poloidal velocity, $v_{\phi}$ is the rotation velocity, and $R_{\text{outflow}}$ is the outflow radius.
For each molecule, the individual $r_0$ values derived from each transverse PV diagram were averaged to obtain the final outer launching radius. The uncertainty was taken as either the standard deviation among the individual $r_0$ values or the propagated error, whichever was larger.

\subsection{Calculation of disk wind mass-loss rate and stellar mass accretion rate}\label{sec:mass_accVSmass_wind}
The mass-loss rate of the disk wind can be calculated from the derived flow parameters using the spectrum extracted from the disk wind region and molecular transition information that traces the disk wind. Following \citet{Nazari2024}, we estimated the mass-loss rate of the disk wind traced by H$_{2}$CO at each slice using the equation:
\begin{align}\label{eq:m_W}
\dot M_{\text{DW}}=1.4m_{\text{p}}(N_{\mathrm{H_{2}CO}}dr)v_{\text{z}}/X_{\mathrm{H_{2}CO}},
\end{align}
where $m_{\text{p}}$ is the proton mass, $X_{\mathrm{H_{2}CO}}$ is the abundance of H$_{2}$CO relative to hydrogen, $dr$ is the length used to extract the spectrum, $N_{\mathrm{H_{2}CO}}$ is the column density of H$_{2}$CO, and $v_{\text{z}}$ is the axial velocity component of the outflow at a given slice. The factor of 1.4 accounts for the contribution of helium to the total gas mass, assuming a helium abundance of 10$\%$ by number. 
\par
\begin{figure*}[!htp]
 \centering{
    \includegraphics[width=0.9\textwidth]{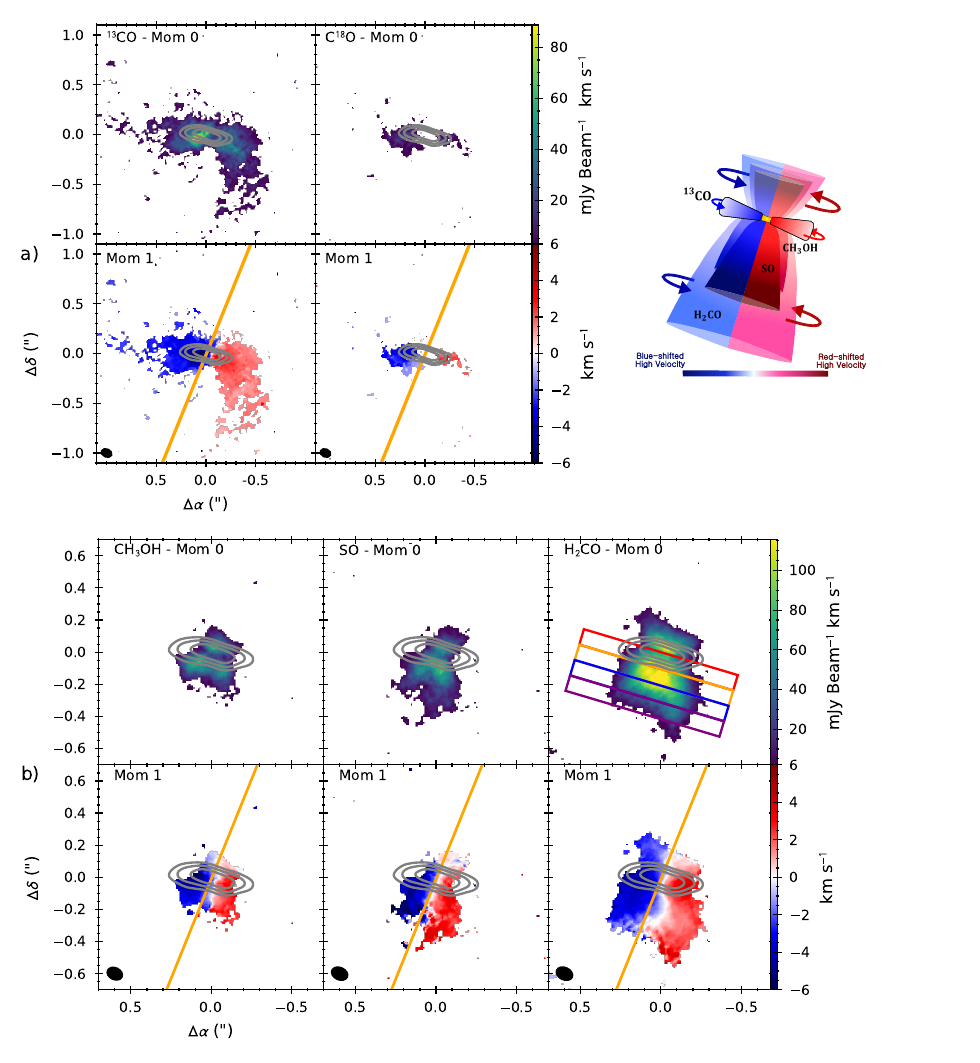}
            }
\caption{{\bf Integrated intensity and velocity maps of molecular lines tracing disk and outflows.}
{\bf a} Integrated intensity (moment 0) and intensity-weighted velocity (moment 1) maps of molecular lines tracing the disk, together with a schematic illustration of the disk and outflow morphology in HOPS~358. {\bf b} Moment 0 and moment 1 maps of molecular lines tracing outflows. The moment maps are constructed using only emission above 4$\sigma$$_{\text{mole}}$ within a velocity range of -11 to 11 \kms\, with respect to systemic velocity, $v_{\text{sys}}$ = 11 \kms. 
The gray contours in each panel represent the 1.3~mm continuum emission at 10, 20, and 40$\sigma$$_{\text{cont}}$.
Details of $\sigma_{\text{mole}}$ and $\sigma_{\text{cont}}$ are provided in Methods, subsection~ALMA observations.
The synthesized beam size corresponding to each molecular dataset is indicated by a black ellipse in the lower left corner of each moment 1 map. The illustration in the upper right presents a schematic view of a rotating disk and disk wind-driven outflows, retaining the same rotation feature as the disk. The orange solid line at each moment 1 map represents the jet/outflow axis (approximately 158\degree). Colored rectangles in the moment 0 map of H$_2$CO indicate slices---centered at 0.05\arcsec, 0.15\arcsec, 0.25\arcsec, and 0.35\arcsec\, from the disk mid-plane, each with a width of 0.1\arcsec---that are used to construct the corresponding transverse PV diagrams in Figure~\ref{fig:fig2}.}
\label{fig:fig1}
\end{figure*}
\begin{figure*}[!htp]
 \centering{
    \includegraphics[width=0.9\textwidth]{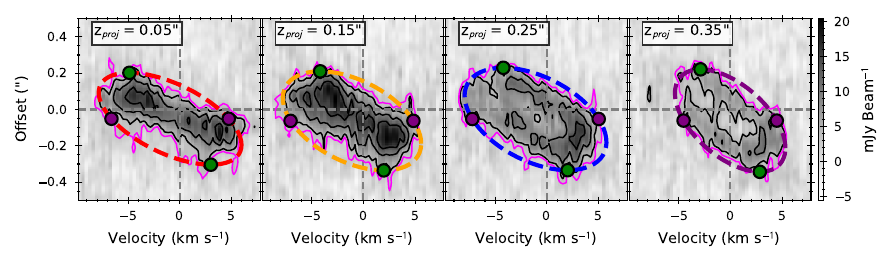}
            }
\caption{{\bf Transverse position-velocity diagrams of H$_2$CO emission across the outflow axis.}
Each slice position corresponding to the transverse PV diagrams is shown in the moment 0 map of H$_2$CO in Figure~\ref{fig:fig1}. The value of z$_{\text{proj}}$ shown on each PV diagram indicates the central position of the respective slice. The colored dashed ellipse on each PV diagram represents the best-fit result from elliptical fitting. The purple and green dots correspond to the same reference points as those shown in Supplementary Fig.~\ref{extfig:SupplyFig1}. Black contour levels correspond to 3, 6, and 9$\sigma$$_{\text{H$_{2}$CO}}$, while the magenta contour level corresponds to 2$\sigma$$_{\text{H$_{2}$CO}}$.}
\label{fig:fig2}
\end{figure*}

In order to obtain the column density of H$_{2}$CO, we first extracted the averaged spectrum within the disk wind region for each slice (see each colored rectangle in Figure~\ref{fig:fig1}). Each spectrum was then fitted with a double-Gaussian profile since the clear rotational feature of the disk wind necessitated modeling with two components to accurately determine the integrated intensity.
\par
\begin{figure*}[!htp]
 \centering{
    \includegraphics[width=0.9\textwidth]{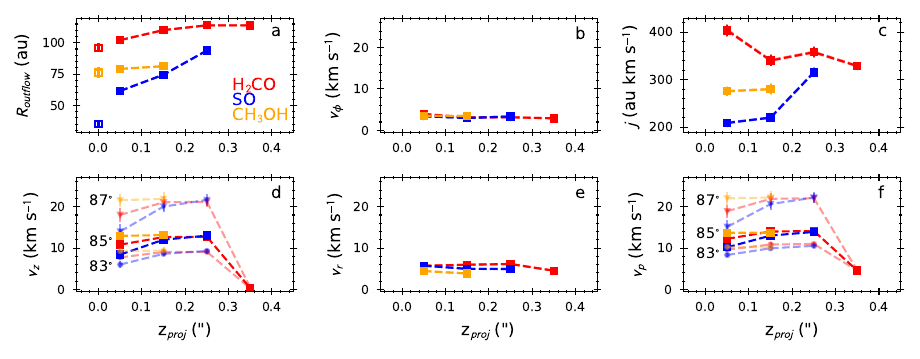}
            }
\caption{{\bf Outflow properties derived from the PV diagrams of three molecular lines.}
{\bf a} Outflow radius. The open symbols represent the outflow radius at z$_{\mathrm{proj}}$ = 0, $R_{0,\mathrm{geom}}$, estimated from parabolic fitting to the measured outflow radii for each molecule. The associated error bars indicate the 1$\sigma$ uncertainties of the $R_{0,\mathrm{geom}}$ derived from the covariance matrix of the parabolic fit. {\bf b} Rotation velocity. {\bf c} Specific angular momentum, $j = R_{\text{outflow}}\times v_{\phi}$. {\bf d} Axial velocity. {\bf e} Radial expansion velocity. {\bf f} Poloidal (or outflowing) velocity. The red, orange, and blue markers represent the outflow properties derived from the H$_{2}$CO,  CH$_{3}$OH, and SO molecular lines, respectively. Error bars represent the 1$\sigma$ statistical uncertainties propagated from the bootstrap-derived ellipse-fitting parameters. For the axial and poloidal velocities, which are particularly sensitive to the nearly edge-on geometry, the effect of inclination is illustrated by showing values computed for $i = 83^\circ$ and $87^\circ$, indicated by circle and triangle symbols, respectively. These values bracket the estimated inclination uncertainty of $85^{+2}_{-2}{^\circ}$.}
\label{fig:fig3}
\end{figure*}
\begin{figure*}[!htp]
 \centering{
    \includegraphics[width=0.9\textwidth]{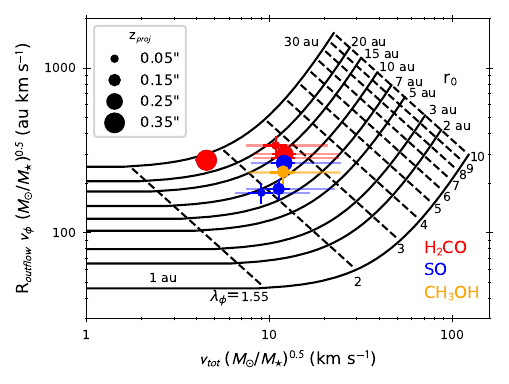}
            }
\caption{{\bf Comparison of observed outflow kinematics with theoretical MHD disk wind models.} 
The outflow properties are normalized by the square root of the protostellar mass. The color scheme is the same as in Figure~\ref{fig:fig3}; the red, orange, and blue symbols represent values derived from H$_{2}$CO, CH$_{3}$OH, and SO, respectively. Each symbol indicates the value calculated from the transverse PV diagram at the corresponding projected distance along the outflow axis from the disk, z$_{\text{proj}}$, assuming $i=85\degree$. Error bars represent the 1$\sigma$ statistical uncertainties of the derived outflow properties, propagated from the bootstrap-derived ellipse-fitting parameters and including the inclination correction and mass normalization.
The semi-transparent error bars indicate values recalculated assuming $i=83\degree$ (left) and $i=87\degree$ (right), illustrating the systematic effect of the inclination uncertainty.
The curves represent the expected relations from steady self-similar MHD disk wind models. Solid curves correspond to different launching radii, $r_0$. Dashed curves correspond to different magnetic lever arm parameters, $\lambda_{\phi}$, which provide lower limits on the theoretical wind lever arm parameter.}
\label{fig:fig4}
\end{figure*}
\begin{figure*}[!htp]
 \centering{
    \includegraphics[width=0.9\textwidth]{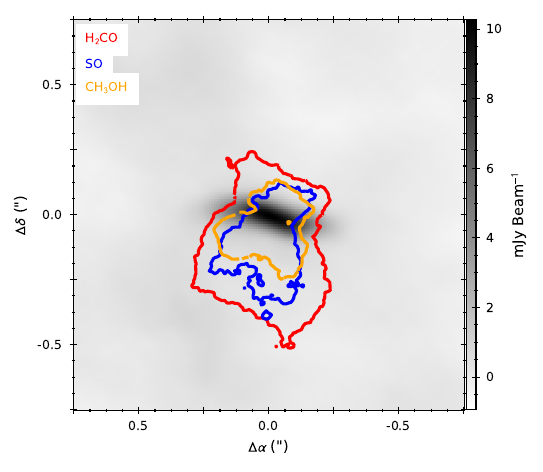}
            }
\caption{{\bf Spatial distribution of outflow-tracing molecular emission.} 
The background image shows the dust continuum. The red contour represents the outer boundary of the moment 0 map of H$_{2}$CO emission, while the blue and orange contours show the moment 0 maps of SO and CH$_{3}$OH, respectively. Contour levels for each molecule correspond to 8.7 mJy Beam$^{-1}$ km s$^{-1}$ for H$_{2}$CO, 11.9 mJy Beam$^{-1}$ km s$^{-1}$ for SO, and 11.1 mJy Beam$^{-1}$ km s$^{-1}$ for CH$_{3}$OH, which represent two times the root mean square noise level of each moment 0 map.}
\label{fig:fig5}
\end{figure*}

We assume that the disk wind is optically thin and in local thermal equilibrium (LTE) conditions. The excitation temperature was fixed at 22.4 K, corresponding to the peak of the averaged spectrum extracted over the entire disk wind region. These assumptions imply that the derived $N_{\mathrm{H_{2}CO}}$ and the resulting mass-loss rate represent lower limits.
The $dr$ was estimated by dividing the area used to extract each spectrum by 0.1\arcsec, which corresponds to the width of each slice. The $v_{\text{z}}$ associated with $N_{\mathrm{H_{2}CO}}$ for each slice is shown in Figure~\ref{fig:fig3}. 
\par
The abundance of H$_{2}$CO, $X_{\text{H$_{2}$CO}}$, has been reported for outflow regions of several sources \citep{Bachiller1997, Jorgensen2004, Tafalla2010}. In this study, we adopted a value of 3$\times$10$^{-7}$, as measured in the L1157 outflow \citep{Bachiller1997, Jorgensen2004} since L1157 is a Class 0 protostar with a chemically rich outflow, similar to HOPS 358. Finally, the total mass-loss rate is estimated by first calculating the average mass-loss rate from all slices and then multiplying it by two to account for the fact that the average value represents only one side of the outflow.
\par
The mass accretion rate is calculated by assuming that the majority of the liberated gravitationally energy due to the infalling matter onto the star is radiated away:
\begin{align}\label{eq:m_acc}
\dot M_{\text{acc}}=\frac{L_{\text{acc}}R_{\star}}{f_{\text{acc}}GM_{\star}},
\end{align}
where $L_{\text{acc}}$ is accretion luminosity, $R_{\star}$ is protostar radius, $f_{\text{acc}}$ is the fraction of the accretion energy that is radiated, ranging from 0 to 1 \citep{hartmann2016}. \citet{Furlan2016} performed spectral energy distribution fitting for HOPS 358 and derived a total luminosity of 60.4 $L_{\odot}$. Since accretion is the primary source of total luminosity in the Class 0 stage, we assumed that the total luminosity corresponds to the accretion luminosity. We adopted a protostar radius of 2 $R_{\odot}$, a typical value during the Class 0 phase \citep{cfLee2020, Podio2021}. The $f_{\text{acc}}$ remained uncertain and we assumed $f_{\text{acc}}$ = 1, which constrains the mass accretion rate as a lower limit.
\clearpage

\backmatter

\vspace{0.3cm}
\noindent
{\bf Data availability}\\
This paper makes use of the following ALMA data from project ADS/JAO.ALMA\#2023.1.01245.S, which are publicly available from the ALMA Science Archive (https://almascience.nao.ac.jp/aq/). Source Data are provided with this paper. Additional data products supporting the findings of this study are available at Zenodo (\url{https://zenodo.org/records/18960708}) \cite{ch0720_2026_18960708}. The reduced ALMA cube data are available from the corresponding author upon request.

\vspace{0.3cm}
\noindent
{\bf Code availability}\\
The analysis codes used to generate the results and figures in this study are available at Zenodo (\url{https://zenodo.org/records/18960708}) \cite{ch0720_2026_18960708}. The analysis was performed using Python with scientific libraries, as well as CASA\citep{McMullin2007} for ALMA data reduction. All software packages used are publicly available.

\bibliography{ref}




\bmhead{Acknowledgements}
This work was supported by the New Faculty Startup Fund from Seoul National University (J.-E.L.) and the National Research Foundation of Korea (NRF) grant funded by the Korean government (MSIT; grant number 2021R1A2C1011718 and RS-2024-00416859 to J.-E.L. and C.-H.K.). D.J. is supported by NRC Canada and by an NSERC Discovery Grant. G.J.H. is supported by grant IS23020 from the Beijing Natural Science Foundation. C.-F. L. acknowledges a grant from the National Science and Technology Council of Taiwan (112-2112- M-001-039-MY3).

This paper makes use of the following ALMA data: ADS/JAO.ALMA\#2023.1.01245.S. ALMA is a partnership of ESO (representing its member states), NSF (USA), and NINS (Japan), together with NRC (Canada), MOST and ASIAA (Taiwan), and KASI (Republic of Korea), in cooperation with the Republic of Chile. The Joint ALMA Observatory is operated by ESO, AUI/NRAO, and NAOJ. The National Radio Astronomy Observatory is a facility of the U.S. National Science Foundation operated under cooperative agreement by Associated Universities, Inc.

\vspace{0.3cm}
\noindent
{\bf Author Contribution}\\
C.-H.K. led the ALMA Cycle 10 program (Project ID: 2023.1.01245.S). J.-E.L., D.J., G.J.H., C.-F.L., and L.F. participated in the ALMA proposal. C.-H. K. performed the reduction and analysis of the ALMA data.  J.-E.L. and C.-H.K. planned the ALMA observations and wrote the manuscript. C.-H.K., J.-E.L., D.J., G.J.H., C.-F.L., L.F., and P.D.S discussed the results and/or commented on the manuscript.

\vspace{0.3cm}
\noindent
{\bf Conflict of interest}\\
The authors declare no competing interests.

\clearpage

\vspace{1cm}

\vspace{1cm}

\vspace{1cm}

\vspace{1cm}

\vspace{1cm}



\clearpage

\begin{appendices}
\title{{\bf Supplementary Information File}}
\maketitle

\setcounter{table}{0}
\renewcommand{\thetable}{\arabic{table}}
\renewcommand{\tablename}{Supplementary Table}

\begin{table}[htp]
\caption{\textbf{SLAM results accounting for the uncertainties in the inclination angle ($i$) and the position angle (P.A.) of the disk major axis.} 
For each combination of inclination and P.A., the values derived using the ridge and edge methods are listed separately. Each entry is given in the form $M_{\star} \pm \sigma_{M_{\star}}$ / $p \pm \sigma_{p}$, where $M_{\star}$ is the protostar mass and $p$ is the power-law index of the velocity profile ($v \propto r^{-p}$).}\label{<tab:protostar_mass>}
\label{tab:protostar_mass}

\centering
\begin{tabular}{cccc}
\toprule
$M_{\star}^{\dagger}$ ($M_{\odot}$) / $p^{\ddagger}$ &
P.A. of 253.3\degree & P.A. of 253.4\degree & P.A. of 253.6\degree \\
\midrule
\vspace{0.3cm}
$i$ = 83\degree & (ridge)
1.01$\pm$0.02 / 0.53$\pm$0.01 & 
1.02$\pm$0.02 / 0.53$\pm$0.01 & 
1.06$\pm$0.02 / 0.56$\pm$0.01 \\
 & (edge) 1.83$\pm$0.03 / 0.71$\pm$0.02 & 
 1.83$\pm$0.03 / 0.71$\pm$0.02 & 
 1.81$\pm$0.03 / 0.70$\pm$0.02 \\ 
\\
\vspace{0.3cm}
$i$ = 85\degree & (ridge)
1.01$\pm$0.02 / 0.53$\pm$0.01 & 
1.01$\pm$0.02 / 0.53$\pm$0.01 & 
1.05$\pm$0.02 / 0.56$\pm$0.01 \\
 & (edge) 1.82$\pm$0.03 / 0.71$\pm$0.02 & 
 1.82$\pm$0.03 / 0.71$\pm$0.02 & 
 1.80$\pm$0.03 / 0.70 $\pm$0.02 \\ 
\\
\vspace{0.3cm}
$i$ = 87\degree & (ridge)
1.00$\pm$0.02 / 0.53$\pm$0.01 & 
1.00$\pm$0.02 / 0.53$\pm$0.01 & 
1.04$\pm$0.02 / 0.56$\pm$0.01 \\
 & (edge) 1.81$\pm$0.03 / 0.71$\pm$0.02 & 
 1.81$\pm$0.03 / 0.71$\pm$0.02 & 
 1.79$\pm$0.03 / 0.70$\pm$0.02 \\ 
\bottomrule
\end{tabular}

\vspace{0.3em}
\begin{flushleft}
\footnotesize
\textdagger\ Protostar mass.\\
$\ddagger$\ Power-law index of the velocity profile ($v \propto r^{-p}$).
\end{flushleft}

\end{table}

\clearpage

\counterwithin{figure}{section}
\renewcommand{\thesection}{Supplementary}
\renewcommand{\thefigure}{\arabic{figure}}
\renewcommand{\figurename}{Supplementary Fig.}

\begin{figure*}[!htp]
 \centering{
    \includegraphics[width=0.9\textwidth]{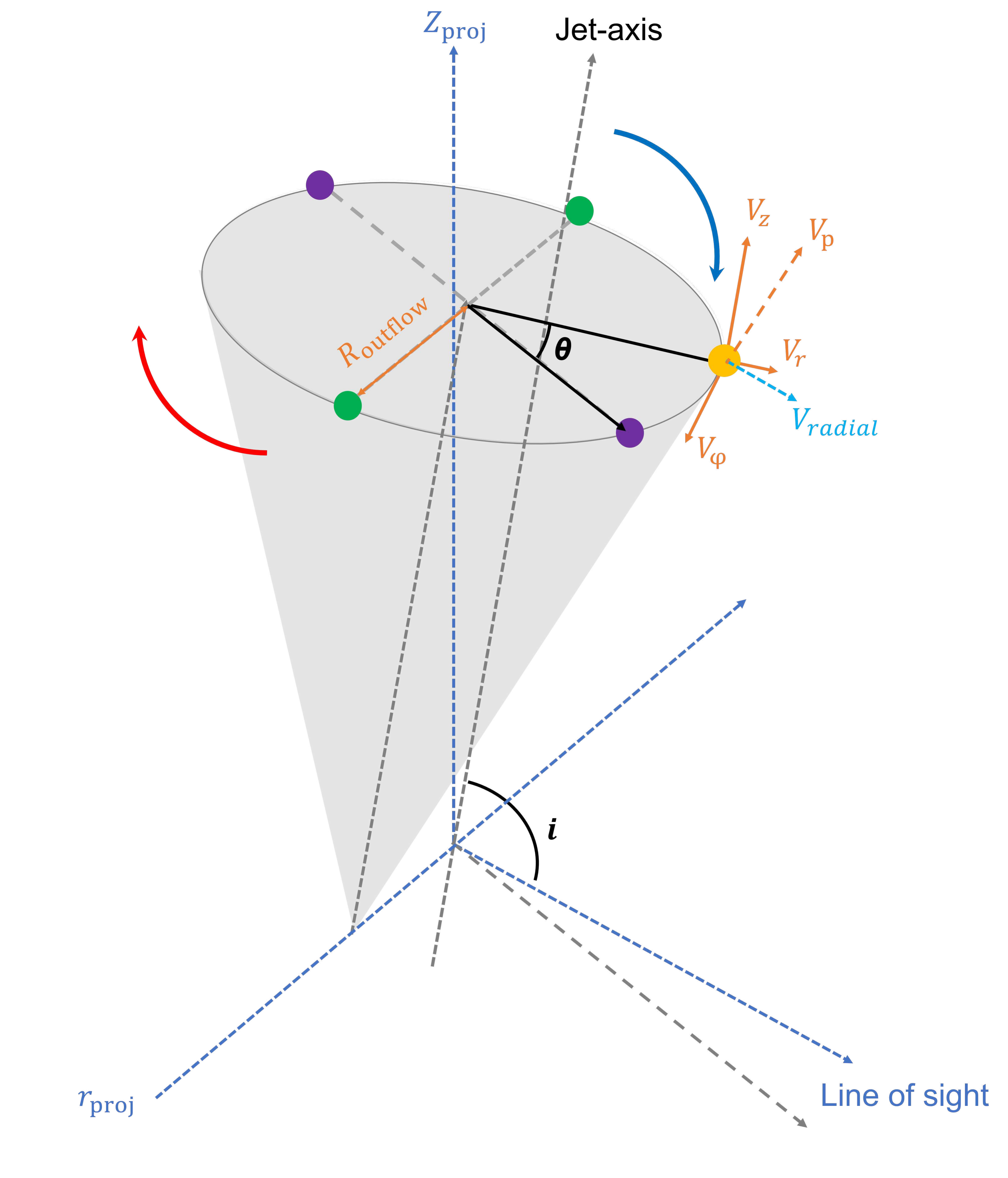}
            }
\caption{{\bf Schematic view of the outflow assuming a thin-shell structure.} 
The grey conical surface represents a geometrically thin outflow shell surrounding the jet axis. The projected coordinate axes, $z_{\text{proj}}$ and $r_{\text{proj}}$, lie in the plane of the sky and correspond to the projected jet axis and its perpendicular direction, respectively. The jet axis is inclined by an angle $i$ with respect to the line of sight. An example gas parcel located on the shell is marked by an orange dot and shown together with its velocity components, axial velocity, $v_{\text{z}}$, radial expansion velocity, $v_{\text{r}}$, and rotation velocity, $v_{\phi}$. The poloidal velocity, $v_{\text{p}}$, represents the outflowing velocity along the wind streamline. The observed line-of-sight velocity, $v_{\text{radial}}$, corresponds to the projection of the local velocity vector onto the line of sight. The azimuthal angle, $\theta$, is defined in the outflow shell cross-section. Purple dots indicate gas parcels at $\theta$ = 0 and $\theta=\pi$, while green dots mark parcels at $\theta$ = $\pm\frac{\pi}{2}$, which are used to derive the velocity components and outflow radius from the transverse PV diagrams.}
\label{extfig:SupplyFig1}
\end{figure*}

\begin{figure*}[!htp]
\centering{
    \includegraphics[width=0.95\textwidth]{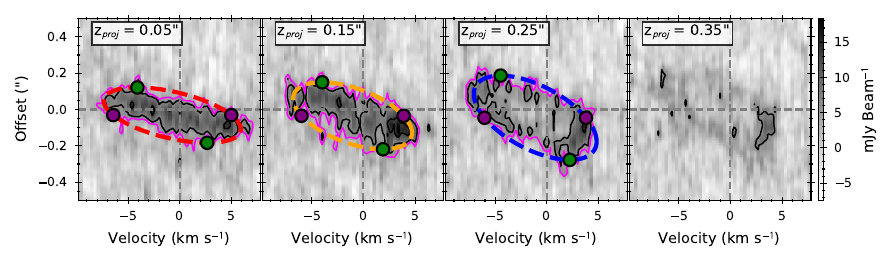}
            }
\caption{{\bf Transverse position-velocity (PV) diagrams of SO emission perpendicular to the outflow axis at the corresponding slices}. The value of z$_{\text{proj}}$ shown on each PV diagram indicates the center position of each slice. The colored dashed ellipse on each PV diagram shows the best-fit result from elliptical fitting. The purple and green dots correspond to the same reference points as those shown in Supplementary Fig.~\ref{extfig:SupplyFig1}. Black contour levels correspond to 3, 6, and 9$\sigma$$_{\text{SO}}$, while the magenta contour level corresponds to 2$\sigma$$_{\text{SO}}$.}
\label{extfig:SupplyFig2}
\end{figure*}

 \begin{figure*}[!htp]
 \centering{
    \includegraphics[width=0.95\textwidth]{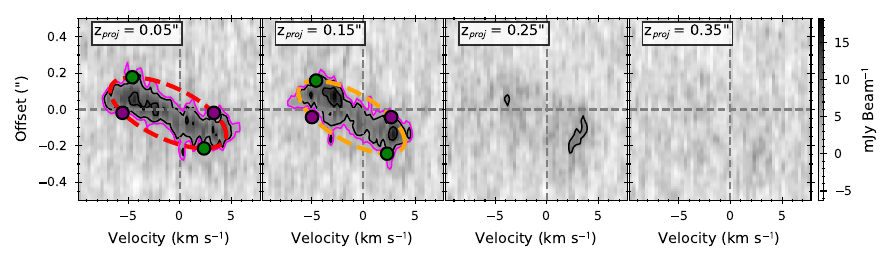}
            }
\caption{{\bf Transverse position-velocity (PV) diagrams of CH$_{3}$OH emission perpendicular to the outflow axis at the corresponding slices.} The value of z$_{\text{proj}}$ shown on each PV diagram indicates the center position of each slice. The colored dashed ellipse on each PV diagram shows the best-fit result from elliptical fitting. The purple and green dots correspond to the same reference points as those shown in Supplementary Fig.~\ref{extfig:SupplyFig1}. Black contour levels correspond to 3, 6, and 9$\sigma$$_{\text{CH$_{3}$OH}}$, while the magenta contour level corresponds to 2$\sigma$$_{\text{CH$_{3}$OH}}$.}
\label{extfig:SupplyFig3}
\end{figure*}

 \begin{figure*}[!htp]
 \centering{
    \includegraphics[width=0.9\textwidth]{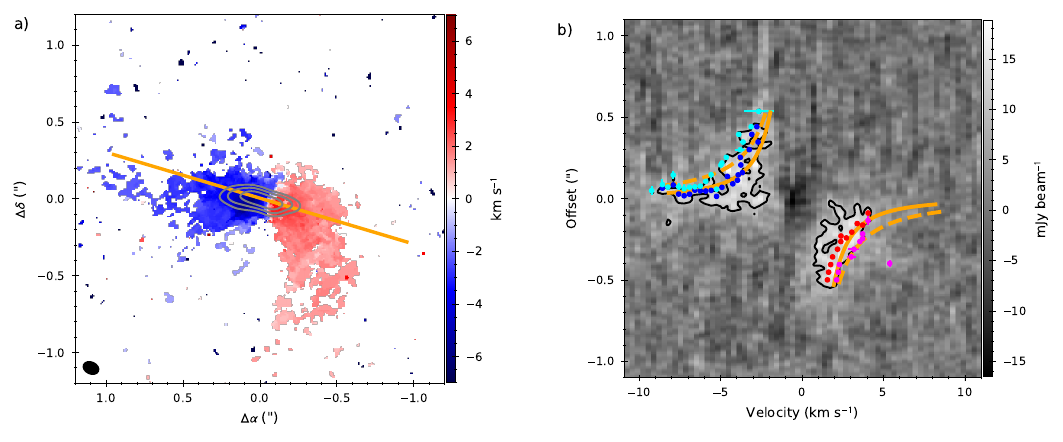}
            }
\caption{{\bf PV diagram along the disk major axis for $^{13}$CO.}
{\bf a}: Moment 1 map of $^{13}$CO, identical to Figure~\ref{fig:fig1}. The orange solid line indicates the PV cut corresponding to the disk major axis (254.3\degree). {\bf b}: PV diagram of $^{13}$CO along the disk major axis. The cyan and magenta symbols mark the points extracted using the edge method, corresponding to the blue- and red-shifted sides, respectively. The blue and red symbols denote the points extracted using the ridge method on the blue- and red-shifted sides, respectively.
The orange dashed and solid lines show the best-fit results from the edge and ridge methods, respectively. The black contours correspond to the 4$\sigma_{\mathrm{^{13}CO}}$ level.}
\label{extfig:SupplyFig4}
\end{figure*}





\end{appendices}
\end{document}